\documentclass[aps,showpacs,showkeys,pre,twocolumn,superscriptaddress,floatfix]{revtex4}
\usepackage{graphicx,bm,amssymb,amsmath,dcolumn,hyperref}
\usepackage{multirow}
\usepackage{enumitem}
\usepackage{color}
\usepackage{amsfonts}
\usepackage{amsthm}

\begin{document}
\newtheorem{lemma}{Lemma}
\newcommand{\be}{\begin{equation}}
\newcommand{\ee}{\end{equation}}
\newcommand{\ind}{\mathbf{1}}
\renewcommand{\Pr}{\mathbb{P}}
\newcommand{\ZZ}{\mathbb{Z}}
\newcommand{\EE}{\mathbb{E}}
\newcommand{\LL}{\mathbb{L}}
\newcommand{\fL}{{\mathfrak L}}
\newcommand{\scrA}{{\mathcal A}}
\newcommand{\scrB}{{\mathcal B}}
\newcommand{\scrL}{{\mathcal L}}
\newcommand{\scrN}{{\mathcal N}}
\newcommand{\scrS}{{\mathcal S}} 
\newcommand{\scrs}{{\mathcal s}}
\newcommand{\scrP}{{\mathcal P}}
\newcommand{\scrM}{{\mathcal M}}
\newcommand{\scrO}{{\mathcal O}}
\newcommand{\scrR}{{\mathcal R}}
\newcommand{\scrC}{{\mathcal C}}
\newcommand{\scrl}{{\mathcal l}}
\newcommand{\dm}{d_{\rm min}}
\newcommand{\rhojunction}{\rho_{\rm j}}
\newcommand{\rhojunctionLim}{\rho_{{\rm j},0}}
\newcommand{\rhobranch}{\rho_{\rm b}}
\newcommand{\rhobranchLim}{\rho_{{\rm b},0}}
\newcommand{\rhononbridge}{\rho_{\rm n}}
\newcommand{\rhononbridgeLim}{\rho_{{\rm n},0}}
\newcommand{\leafFreeHull}{H_{\rm lf}}
\newcommand{\bridgeFreeHull}{H_{\rm bf}}
\newcommand{\percolationCluster}{C_1}
\newcommand{\leafFreeCluster}{C_{\rm lf}}
\newcommand{\bridgeFreeCluster}{C_{\rm bf}}
\newcommand{\dF}{d_{\rm F}}
\newcommand{\dB}{d_{\rm B}}
\newcommand{\dR}{d_{\rm R}}
\newcommand{\dH}{d_{\rm H}}
\newcommand{\dE}{d_{\rm E}}
\newcommand{\ontop}[2]{\genfrac{}{}{0pt}{}{#1}{#2}}

\newcommand{\blue}[1]{\textcolor{blue}{#1}}
\newcommand{\red}[1]{\textcolor{red}{#1}}
\newcommand{\green}[1]{\textcolor{green}{#1}}

\newcommand{\diagramFig}{\includegraphics[scale=0.25]{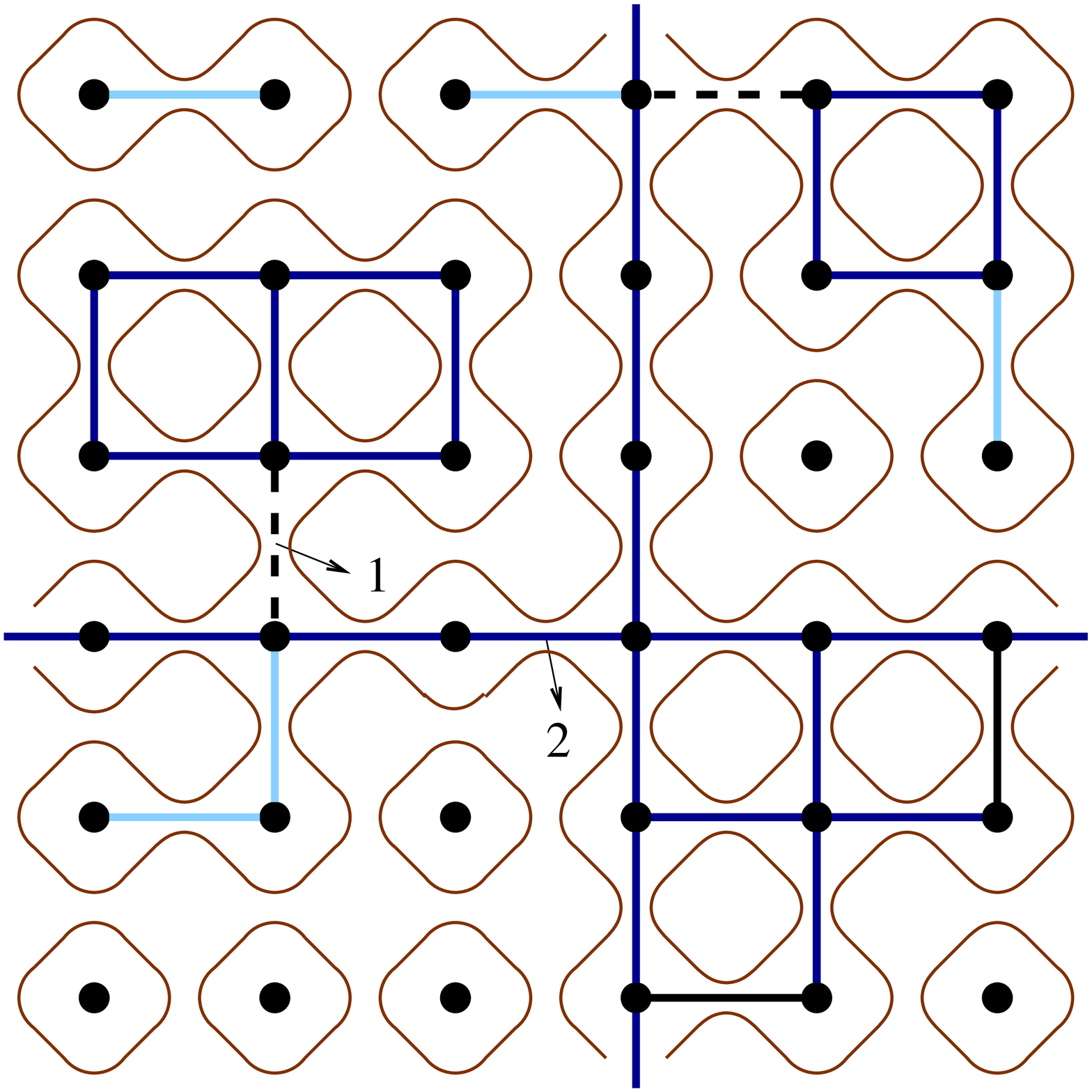}}
\newcommand{\configurationFig}{\includegraphics[trim = 0mm 0mm 0mm 0mm, clip, scale=0.68]{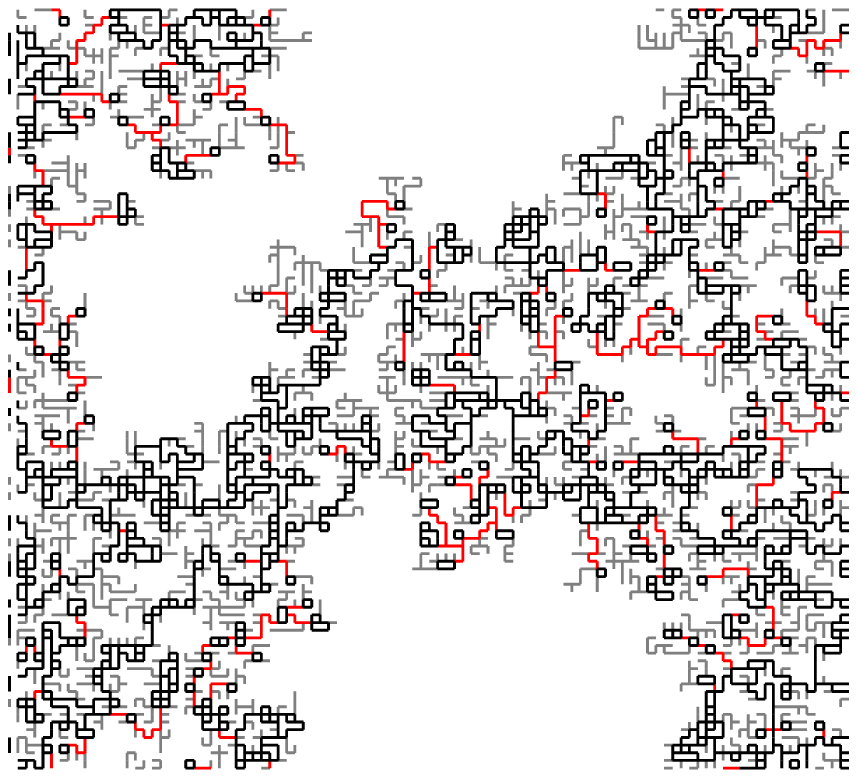}}
\newcommand{\densitiesFig}{\includegraphics[scale=0.7]{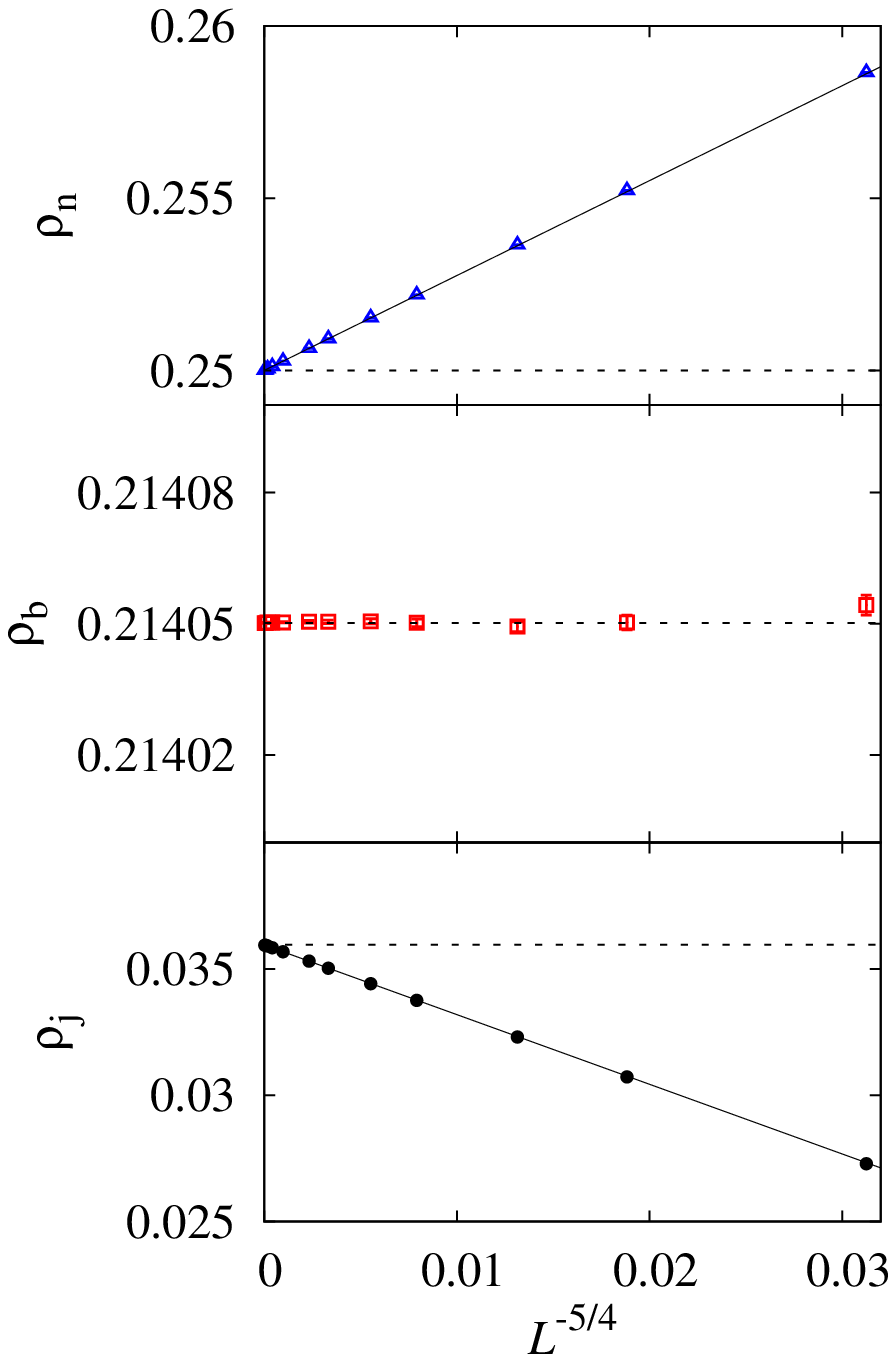}}
\newcommand{\leafFreeFig}{\includegraphics[scale=0.68]{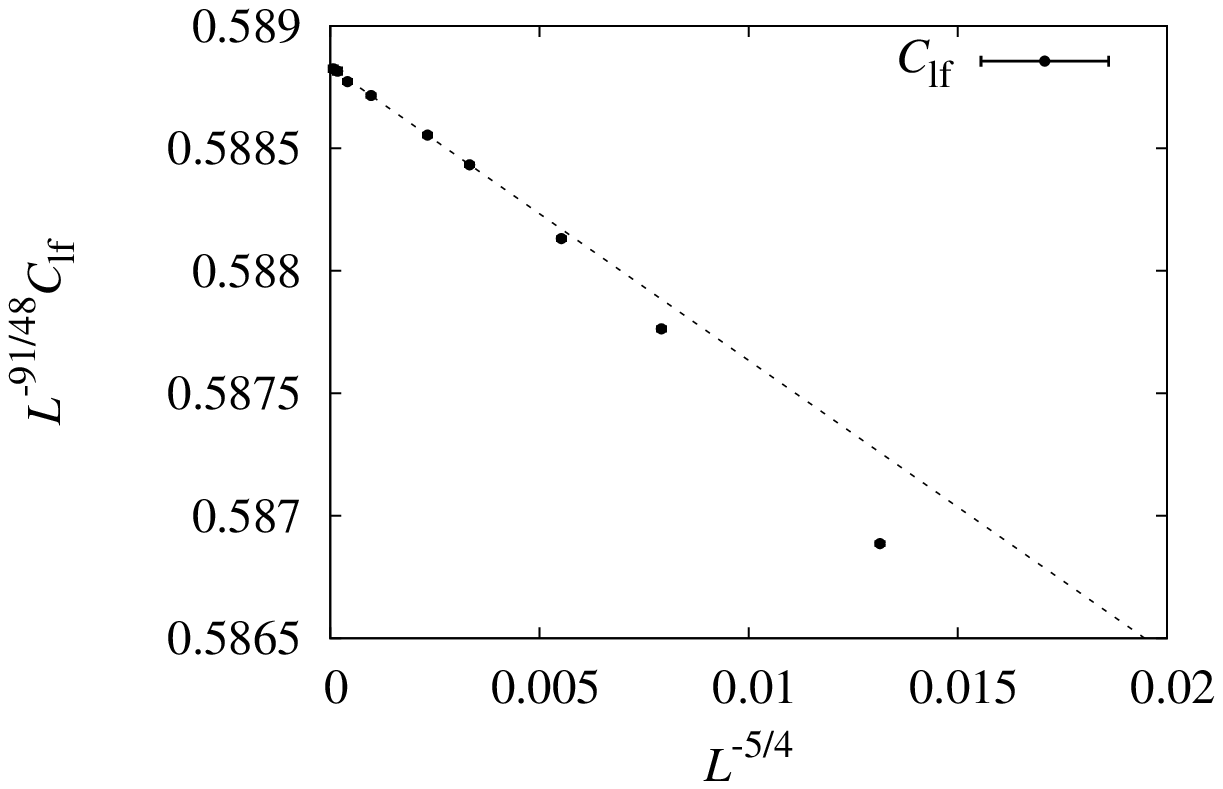}}
\newcommand{\backboneFig}{\includegraphics[scale=0.68]{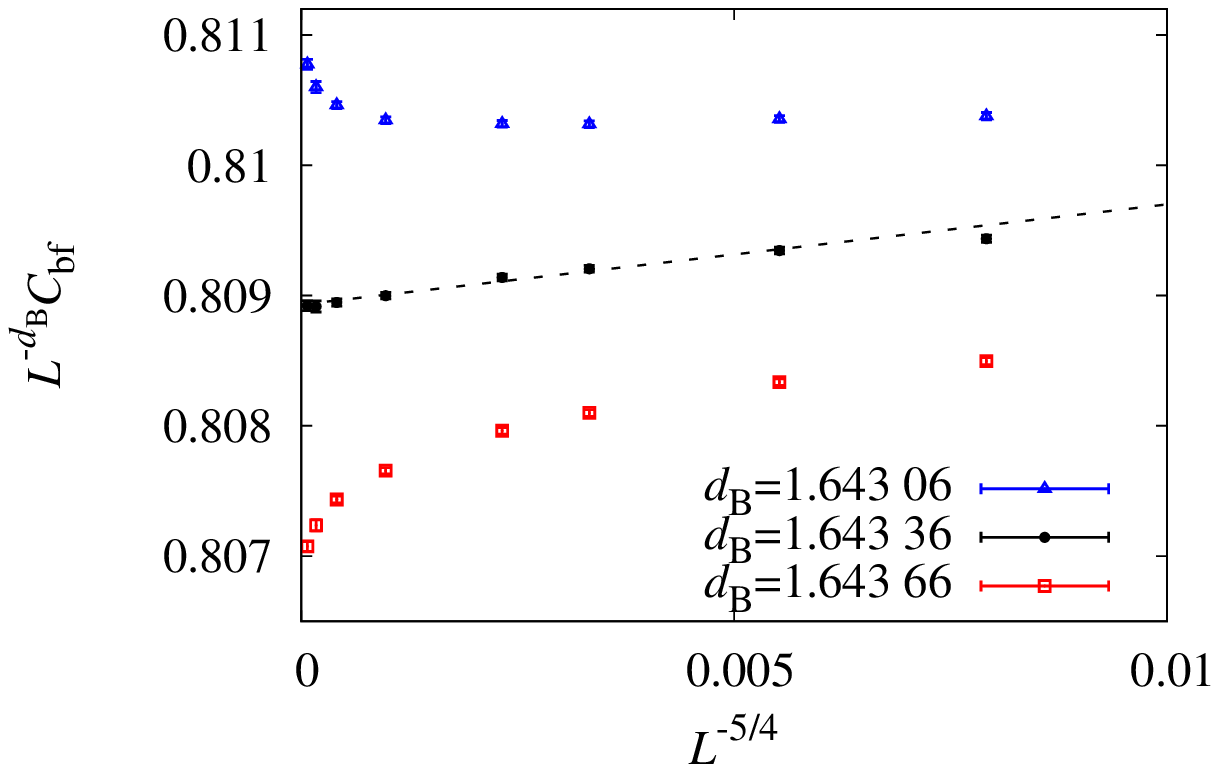}}
\newcommand{\hullFig}{\includegraphics[scale=0.68]{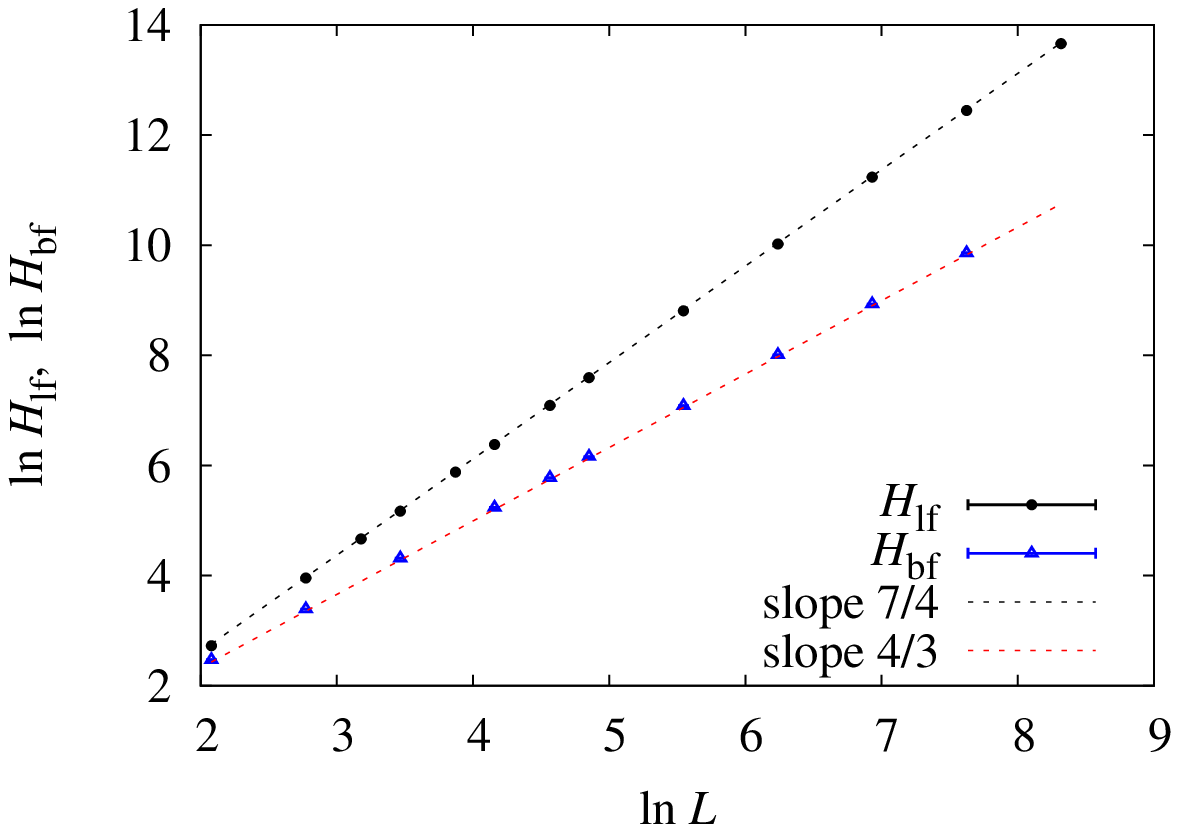}}
\newcommand{\loopFigA}{\includegraphics[trim = 35mm 190mm 40mm 30mm, clip, scale=0.325]{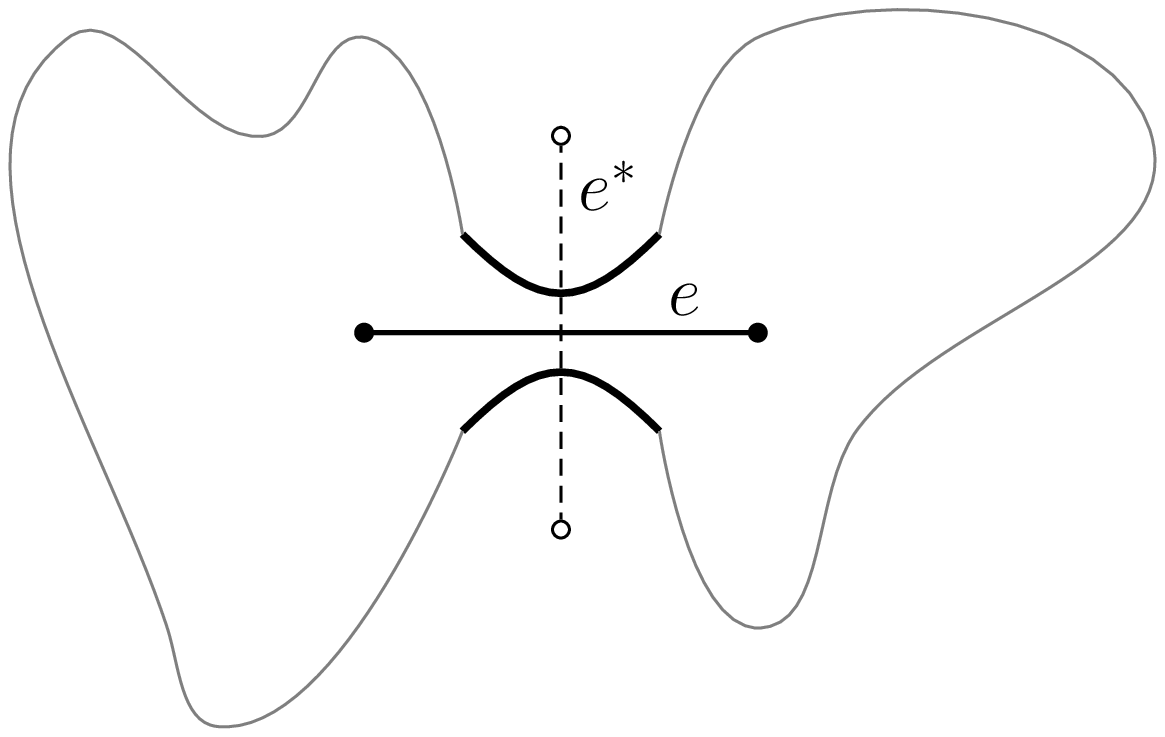}}
\newcommand{\loopFigB}{\includegraphics[trim = 35mm 190mm 40mm 30mm, clip, scale=0.325]{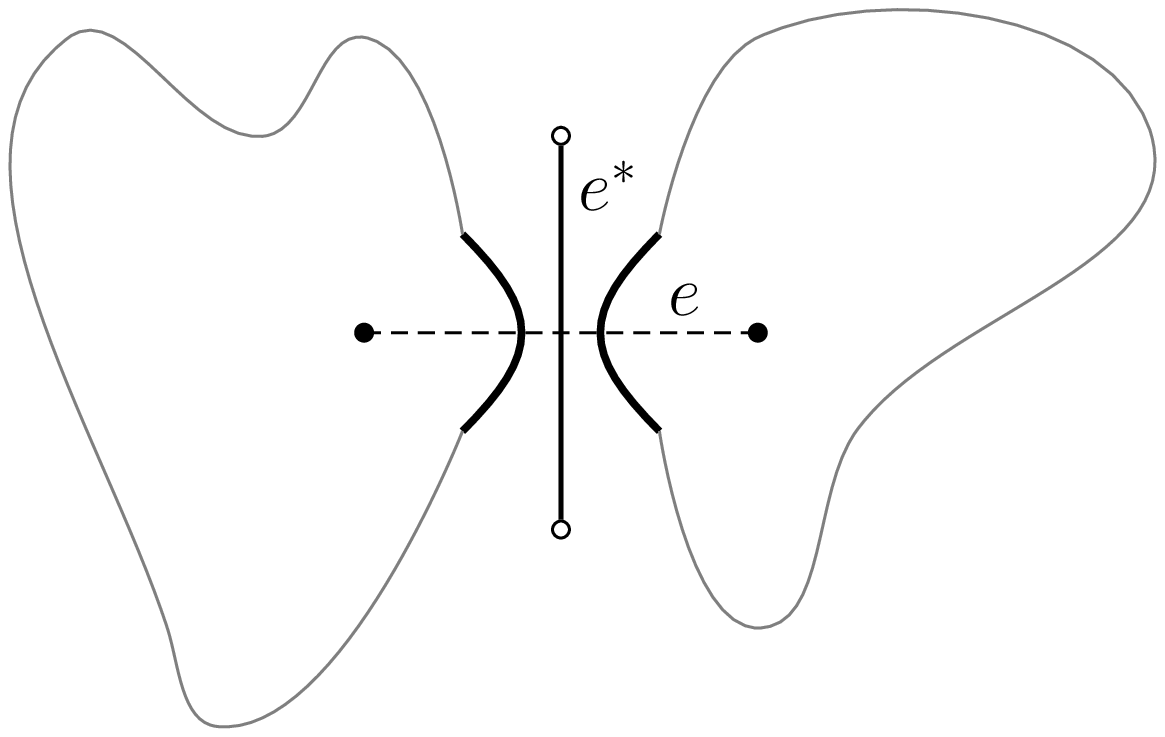}}

\title{Geometric structure of percolation clusters}
\date{\today}
\author{Xiao Xu}
\affiliation{Hefei National Laboratory for Physical Sciences at Microscale and Department of Modern Physics, University of Science and Technology of China, Hefei, Anhui 230026, China}
\author{Junfeng Wang} 
\affiliation{Hefei National Laboratory for Physical Sciences at Microscale and Department of Modern Physics, University of Science and Technology of China, Hefei, Anhui 230026, China}
\author{Zongzheng Zhou}
\email{eric.zhou@monash.edu}
\affiliation{School of Mathematical Sciences, Monash University, Clayton, Victoria~3800, Australia}
\author{Timothy M. Garoni}
\email{tim.garoni@monash.edu}
\affiliation{School of Mathematical Sciences, Monash University, Clayton, Victoria~3800, Australia}
\author{Youjin Deng}
\email{yjdeng@ustc.edu.cn}
\affiliation{Hefei National Laboratory for Physical Sciences at Microscale and Department of Modern Physics, University of Science and Technology of China, Hefei, Anhui 230026, China}

\begin{abstract}
  We investigate the geometric properties of percolation clusters, by studying square-lattice bond percolation on the torus. 
  We show that the density of bridges and nonbridges both tend to 1/4 for large system sizes.
  Using Monte Carlo simulations, we study the probability that a given edge is not a bridge but has both its loop arcs in the same loop, and find that it is governed by the two-arm exponent.
  We then classify bridges into two types: branches and junctions.
  A bridge is a {\em branch} iff at least one of the two clusters produced by its deletion is a tree.
  Starting from a percolation configuration and deleting the branches results in a {\em leaf-free} configuration, while deleting all bridges produces a bridge-free configuration.
  Although branches account for $\approx 43\%$ of all occupied bonds, we find that
  the fractal dimensions of the cluster size and hull length of leaf-free configurations are consistent with those for standard percolation configurations.
  By contrast, we find that the fractal dimensions of the cluster size and hull length of bridge-free configurations are respectively given by the backbone and external perimeter dimensions.
  We estimate the backbone fractal dimension to be $1.643\,36(10)$.
\end{abstract}
\maketitle

\section{Introduction}
\label{Introduction}
One of the main goals of percolation theory~\citep{StaufferAharony03,Grimmett99,BollobasRiordan06} in recent decades has been to understand the geometric structure of percolation clusters.
Considerable insight has been gained by decomposing the incipient infinite cluster into a {\em backbone} plus {\em dangling bonds},
and then further decomposing the backbone into {\em blobs} and {\em red bonds}~\cite{Stanley77}.

To define the backbone, one typically fixes two distant sites in the incipient infinite cluster, 
and defines the backbone to be all those occupied bonds in the cluster which belong to trails~\footnote{A {\em trail} in a graph is sequence of adjacent edges, with no repetitions.}
between the specified sites~\cite{HerrmannStanley84}.
The remaining bonds in the cluster are considered dangling.

Similar definitions apply when considering spanning clusters between two opposing sides of a finite box~\cite{Grassberger92}; this is the so-called {\em busbar} geometry.
The bridges~\footnote{An edge in a graph is a {\em bridge} if its deletion increases the number of connected components.} in the backbone constitute the red bonds, while the remaining bonds define the blobs.
At criticality, the average size of the spanning cluster scales as $L^{\dF}$, with $L$ the linear system size and $\dF$ the fractal dimension.
Similarly, the size of the backbone scales as $L^{\dB}$, and the number of red bonds as $L^{\dR}$. 

While exact values for $\dF$ and $\dR$ are known~\cite{NienhuisJSP84,Coniglio89} (see~\eqref{Eq:exact_exponent}), this is not the case for $\dB$.
In~\cite{AizenmanDuplantierAharony99} however, it was shown that $2-\dB$ coincides with the so-called monochromatic path-crossing exponent ${\hat x}_l^{\scrP}$ with $l=2$.
An exact characterization of ${\hat x}_2^{\scrP}$ in terms of a second-order partial differential equation with specific boundary conditions was given in~\cite{LawlerSchrammWerner02},
for which, unfortunately, no explicit solution is currently known.
The exponent ${\hat x}_2^{\scrP}$ was estimated in~\cite{JacobsenZinnJustin02} using transfer matrices, 
and in~\cite{DengBloteNienhuis04a} by studying a suitable correlation function via Monte Carlo simulations on the torus.

\begin{figure}
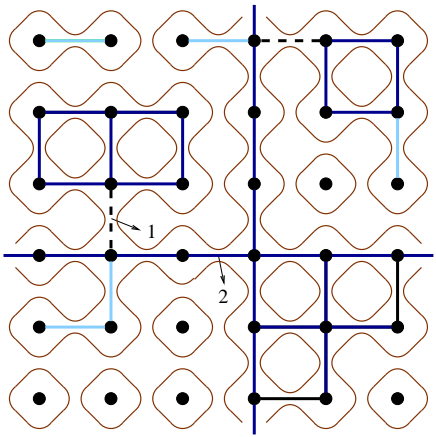

  \diagramFig
  \caption{(Color Online). 
    Decomposition of a percolation configuration into leaf-free and bridge-free configurations. Periodic boundary conditions are applied.
    Nonbridges are denoted by dark blue lines, branches by light blue lines, and junctions by dashed lines.
    The union of the nonbridges and junctions defines the leaf-free configuration.
    Also shown is the BKW loop configuration on the medial lattice, corresponding to the entire percolation configuration.
  }
  \label{Fig:diagram}
\end{figure}

In this paper, we consider a natural partition of the edges of a percolation configuration, and study the fractal dimensions of the resulting clusters.
Specifically, we classify all occupied bonds in a given configuration into three types: branches, junctions and nonbridges.
A bridge is a {\it branch} if and only if at least one of the two clusters produced by its deletion is a tree.
Junctions are those bridges which are not branches.
Deleting branches from percolation configurations produces {\it leaf-free} configurations, and further deleting junctions from leaf-free configurations generates bridge-free configurations.
These definitions are illustrated in Fig.~\ref{Fig:diagram}.

It is often useful to map a bond configuration to its corresponding Baxter-Kelland-Wu (BKW)~\cite{BaxterKellandWu76} loop configuration, as illustrated in Fig.~\ref{Fig:diagram}.
The loop configurations are drawn on the medial graph~\cite{EllisMonaghanMoffatt13}, the vertices of which correspond to the edges of the original graph.
The medial graph of the square lattice is again a square lattice, rotated $45^{\circ}$.
Each unoccupied edge of the original lattice is crossed by precisely two loop arcs, while occupied edges are crossed by none.
The continuum limits of such loops are of central interest in studies of Scharmm L{\"o}wner evolution (SLE)~\cite{KagerNienhuis04,Cardy05}.
At the critical point, the mean length of the largest loop scales as $L^{\dH}$, with $\dH$ the hull fractal dimension.
A related concept is the accessible external perimeter~\cite{GrossmanAharony87}.
This can be defined as the set of sites that have non-zero probability of being visited by a random walker which is initially far from a percolating cluster.
The size of the accessible external perimeter scales as $L^{\dE}$ with $\dE\le\dH$.

In two dimensions, Coulomb-gas arguments~\cite{NienhuisJSP84,SaleurDuplantier87,Coniglio89,Duplantier99} predict the following exact expressions for $\dF$, $\dR$, $\dH$ and $\dE$
\begin{align}
  \dF  &= 2 - (6-g)(g-2)/8g = 91/48\;,  \nonumber  \\
  \dR  &= (4-g)(4+3g)/8g    = 3/4\;,  \nonumber  \\
  \dH  &= 1 + 2/g           = 7/4\;, \nonumber \\
  \dE  &= 1 + g/8           = 4/3\;,
  \label{Eq:exact_exponent}
\end{align}
where for percolation the Coulomb-gas coupling $g=8/3$~\footnote{In terms of the SLE parameter we have $\kappa=16/g=6$.}.
We note that the magnetic exponent $y_h=\dF$, the two-arm exponent~\cite{SaleurDuplantier87} satisfies $x_2 = 2-\dR$, and that for percolation the thermal exponent $y_t=\dR$~\cite{Coniglio82,VasseurJacobsenSaleur12}.
The two-arm exponent gives the asymptotic decay $L^{-x_2}$ of the probability that at least two spanning clusters join inner and outer annuli (of radii O(1) and $L$ respectively) in the plane.
We also note that $\dE$ and $\dH$ are related by the duality transformation $g\mapsto 16/g$~\citep{Duplantier00}.
The most precise numerical estimate for $\dB$ currently known is $\dB = 1.643\,4(2)$~\cite{DengBloteNienhuis04a}.

\begin{figure}
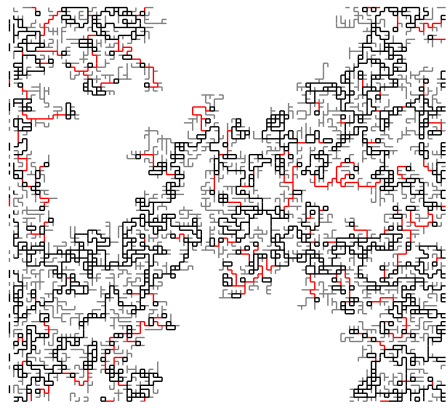

  \configurationFig
  \caption{(Color Online). 
    The largest cluster in a random realization of critical square-lattice bond percolation on an $L\times L$ torus with $L=100$.
    Nonbridges are colored black, branches colored gray, and junctions colored red.
  }
  \label{Fig:configuration}
\end{figure}

We study critical bond percolation on the torus $\ZZ_L^2$, and show that as a consequence of self-duality the density of bridges and nonbridges both tend to 1/4 as $L\to\infty$.
Using Monte Carlo simulations, we observe that despite the fact that around 43\% of all occupied edges are branches, 
the fractal dimension of the leaf-free clusters is simply $\dF$, while their hulls are governed by $\dH$.
By contrast, the fractal dimension of the bridge-free configurations is $\dB$, and that of their hulls is $\dE$.
Fig.~\ref{Fig:configuration} shows a typical realization of the largest cluster in critical square-lattice bond percolation, showing the three different types of bond present.

In more detail, our main findings are summarized as follows.
\begin{enumerate}
\item The leading finite-size correction to the density of nonbridges scales with exponent $-5/4$, consistent with $-x_2$.
  It follows that the probability that a given edge is not a bridge but has both its loop arcs in the same loop decays like $L^{-x_2}$ as $L\to\infty$.
  The leading finite-size correction to the density of junctions also scales with exponent $-5/4$, while the density of branches is almost independent of system size.
\item The fractal dimension of leaf-free clusters is $1.895\,84(4)$, consistent with $\dF=91/48$ for percolation clusters.
\item The hull fractal dimension for leaf-free configurations is $1.749\,96(8)$, consistent with $\dH=7/4$.
\item The fractal dimension for bridge-free clusters is consistent with $\dB$, and we provide the improved estimate $\dB=1.643\,36(10)$.
\item The hull fractal dimension for bridge-free configurations is $1.333\,3(3)$, consistent with $\dE=4/3$.
\end{enumerate}

The remainder of this paper is organized as follows.
Section~\ref{Model_Algorithm_Quantities} introduces the model, algorithm and sampled quantities.
Numerical results are summarized and analyzed in Section~\ref{Results}.
A brief discussion is given in Section~\ref{Discussion}.

\section{Model, Algorithm and Observables}
\label{Model_Algorithm_Quantities}
\subsection{Model}
\label{Model}
We study critical bond percolation on the $L\times L$ square lattice with periodic boundary conditions, with linear system sizes $L=8$, 16, 24, 32, 48, 64, 96, 128, 256, 512, 1024, 2048, and 4096.
To generate a bond configuration, we independently visit each edge on the lattice and randomly place a bond with probability $p=1/2$.
For each system size, we produced at least $7\times 10^{6}$ independent samples; for each $L\le 512$ we produced more than $10^8$ independent samples.

A {\em leaf} in a percolation configuration is a site which is adjacent to precisely one occupied bond.
Given a percolation configuration we generate the corresponding {\em leaf-free} configuration via the following iterative procedure, often referred to as {\em burning}.
For each leaf, we delete its adjacent bond. If this procedure generates new leaves, we repeat it until no leaves remain.
The bonds which are deleted during this iterative process are precisely the branches defined in Section~\ref{Introduction}.

The bridges in the leaf-free configurations are the junctions.
Deleting the junctions from the leaf-free configurations then produces bridge-free configurations.
The algorithm we used to efficiently identify junctions in leaf-free configurations is described in Sec.~\ref{Algorithm}.

\subsection{Algorithm}
\label{Algorithm}
Given an arbitrary graph $G=(V,E)$, the bridges can be identified in $O(|E|)$ time~\cite{Tarjan74,Schmidt13}. 
Rather than applying such graph algorithms to identify the junctions in our leaf-free configurations however, we took advantage of the associated loop configurations. 
These loop configurations were also used to measure the observable $\leafFreeHull$, defined in Section~\ref{Measured quantities}.

Consider an edge $e$ which is occupied in the leaf-free configuration, and denote the leaf-free cluster to which it belongs by $\scrC_e$.
In the planar case, it is clear that $e$ will be a bridge iff the two loop segments associated with it belong to the same loop.
More generally, the same observation holds on the torus provided $\scrC_e$ does not simultaneously wind in both the $x$ and $y$ directions.

If $\scrC_e$ does simultaneously wind in both the $x$ and $y$ directions, loop arguments may still be used, however the situation is more involved.
It clearly remains true that if the two loop segments associated with $e$ belong to different loops, then $e$ is a nonbridge.

Suppose instead that the two loop segments associated with $e$ belong to the same loop, which we denote by $\fL$.
Deleting $e$ breaks $\fL$ into two smaller loops, $\fL_1$ and $\fL_2$.
For each such loop, we let $w_x$ and $w_y$ denote the winding numbers in the $x$ and $y$ directions, respectively, and we define $w = |w_x| + |w_y|$.
As we explain below, the following two statements hold:
\begin{enumerate}[label=(\roman*)]
\item\label{bridge test} If $w(\fL_1)=0$ or $w(\fL_2)=0$, then $e$ is a bridge.
\item\label{nonbridge test} If $w(\fL)=0$ and $w(\fL_1)=1$, then $e$ is a nonbridge.
\end{enumerate}
As an illustration, in Fig.~\ref{Fig:diagram} Edge 1 is a junction while Edge 2 is a nonbridge, despite both of them being bounded by the same loop.
Edge 1 can be correctly classified using statement~\ref{bridge test}, while Edge 2 can be correctly classified using statement~\ref{nonbridge test}.

By making use of these observations, all but very few edges in the leaf-free clusters can be classified as bridges/nonbridges.
We note that in our implementation of the above algorithm, the required $w$ values can be immediately determined from the stored loop configuration without further computational effort.
For the small number of edges to which neither of the above two statements apply,
we simply delete the edge and perform a connectivity check using simultaneous breadth-first search.
This takes $O(L^{\dF-x_2})$ time per edge tested~\cite{DengZhangGaroniSokalSportiello10}.

We now justify statement~\ref{bridge test}.
In this case, the loop $\fL_1$ is contained in a simply-connected region on the surface of the torus.
The cluster contained within the loop $\fL_1$ is therefore disconnected from the remainder of the lattice, implying that $e$ is a bridge.
Edge 1 in Fig.~\ref{Fig:diagram} provides an illustration.

Finally, we justify statement~\ref{nonbridge test}.
In this case, $\fL_1$ and $\fL_2$ either both wind in the $x$ direction, or both in the $y$ direction (one winds in the positive sense, the other in the negative sense).
Suppose they wind in the $y$ direction. It then follows from the definition of the BKW loops that there can be no $x$-windings in the cluster $\scrC_e\setminus e$. 
By assumption however, $\scrC_e$ does contain an $x$-winding, so it must be the case that $e$ belongs to a winding cycle in $\scrC_e$ that winds in the $x$ direction.
The edge $e$ is therefore not a bridge. 
Edge 2 in Fig.~\ref{Fig:diagram} provides an illustration.

\subsection{Measured quantities}
\label{Measured quantities}
From our simulations, we estimated the following quantities.
\begin{enumerate}
\item The mean density of branches $\rhobranch$, junctions $\rhojunction$, and nonbridges $\rhononbridge$. 
\item The mean size of the largest cluster $\percolationCluster$
\item The mean size of the largest leaf-free cluster $\leafFreeCluster$
\item The mean size of the largest bridge-free cluster $\bridgeFreeCluster$
\item The mean length of the largest loop, $\leafFreeHull$, for the loop configuration associated with leaf-free configurations
\item The mean length of the largest loop, $\bridgeFreeHull$, for the loop configuration associated with bridge-free configurations
\end{enumerate}

We note that fewer samples were generated for $\percolationCluster$ and $\bridgeFreeHull$ than for other the quantities.

\section{Results}
\label{Results}
In Sections~\ref{Bond density},~\ref{Fractal dimension of clusters},~\ref{Fractal dimension of loops}, 
we discuss least-squares fits for $\rhobranch$, $\rhojunction$, $\rhononbridge$ and $\leafFreeCluster$, $\bridgeFreeCluster$, $\leafFreeHull$, $\bridgeFreeHull$.
The results are presented in Tables~\ref{Tab:Bond_Density},~\ref{Tab:FD_Clusters} and~\ref{Tab:FD_Loops}.
In Section~\ref{Fitting methodology}, we first make some comments on the ans\"atze and methodology used.

\subsection{Fitting ans\"atze and methodology}
\label{Fitting methodology}
Let $\rho_1$ ($\rho_2$) denote the mean density of occupied edges whose two associated loop segments belong to the same (distinct) loop(s).
From Lemma~\ref{loop lemma} in Appendix~\ref{loop lemma appendix}, we know that for $p=1/2$ bond percolation on $\ZZ_L^2$ we have $\rho_1 = \rho_2 = 1/4$ for all $L$.
In the plane however, an edge is a bridge iff the two associated loop segments belong to the same loop.
We therefore expect that both $\rhononbridge$ and $\rhojunction+\rhobranch$ should converge to 1/4 as $L\to\infty$.

Furthermore, there is a natural interpretation of the quantity $\rhononbridge-\rho_2$.
As noted in Section~\ref{Algorithm}, if the two loop segments associated with an edge belong to different loops, then that edge cannot be a bridge.
This implies that $\rhononbridge-\rho_2$ is equal to the probability of the event that ``a given edge is not a bridge but has both its loop arcs in the same loop''.
Let us denote this event by $\scrB$.
Studying the finite-size behaviour of $\rhononbridge$ will therefore allow us to study the scaling of $\Pr(\scrB)$.
Since $\rhojunction+\rhobranch+\rhononbridge=\rho_1+\rho_2$, it follows that $\rho_1-\rhojunction-\rhobranch$ is also equal to $\Pr(\scrB)$.

Armed with the above observations, we fit our Monte Carlo data for the densities $\rhojunction$, $\rhobranch$ and $\rhononbridge$ to the finite-size scaling ansatz
\begin{equation}
  \rho = \rho_0 + a_1 L^{-y_1} + a_2 L^{-y_2}.
  \label{Eq:Bond_Density}
\end{equation}
We note that since $\rhojunction+\rhobranch+\rhononbridge=1/2$ for all $L$, 
the finite-size corrections of $\rhojunction+\rhobranch$ should be equal in magnitude and opposite in sign to the finite-size corrections of $\rhononbridge$.
Since $\rhononbridge=1/4 + \Pr(\scrB)$, the latter should be positive and the former negative.

Finally, we note that the event $\scrB$ essentially characterizes edges which {\em would} be bridges in the plane, but which are prevented from being bridges on the torus by windings.
By construction, branches always have at least one end attached to a tree, suggesting that they cannot be {\em trapped} in winding cycles in this way. 
This would suggest that it should be $\rhojunction$ that contributes the leading correction of $\rhojunction+\rhobranch$ away from its limiting value of 1/4.

The observables $\leafFreeCluster$, $\bridgeFreeCluster$, $\leafFreeHull$, $\bridgeFreeHull$ are expected to display non-trivial critical scaling, and we fit them to the finite-size scaling ansatz
\begin{equation}
  \scrO = c_0 + L^{d_\scrO}(a_0 + a_1 L^{-y_1} + a_2 L^{-y_2})
  \label{Eq:Clusters_Loops}
\end{equation}
where $d_\scrO$ denotes the appropriate fractal dimension.

As a precaution against correction-to-scaling terms that we failed to include in the fit ansatz, we imposed a lower cutoff $L>L_{\rm min}$ on the data points admitted in the fit,
and we systematically studied the effect on the $\chi^2$ value of increasing $L_{\rm min}$.
Generally, the preferred fit for any given ansatz corresponds to the smallest $L_{\min}$ for which the goodness of fit is reasonable and for which subsequent increases in $L_{\min}$ do 
not cause the $\chi^2$ value to drop by vastly more than one unit per degree of freedom.
In practice, by ``reasonable'' we mean that $\chi^2/\mathrm{DF}\lessapprox 1$, where DF is the number of degrees of freedom.

In all the fits reported below we fixed $y_2=2$, which corresponds to the exact value of the sub-leading thermal exponent~\cite{NienhuisJSP84}.

\subsection{Bond densities}
\label{Bond density}
Leaving $y_1$ free in the fits of $\rhojunction$ and $\rhononbridge$ we estimate $y_1 = 1.250\,5(10)$.
We therefore conjecture that $y_1 = 5/4$, which we note is precisely equal to the two-arm exponent $x_2=5/4$. We comment on this observation further in Section~\ref{Discussion}.

For $\rhobranch$ by contrast, we were unable to obtain stable fits with $y_1$ free.
Fixing $y_1=5/4$, the resulting fits produce estimates of $a_1$ that are consistent with zero.
In fact, we find $\rhobranch$ is consistent with 0.214\,050\,18 for all $L\geq 24$. 
This weak finite-size dependence of $\rhobranch$ is in good agreement with the arguments presented in Section~\ref{Fitting methodology}.

All the fits for $\rhobranch$, $\rhojunction$ and $\rhononbridge$ gave estimates of $a_2$ consistent with zero. 
We therefore set $a_2=0$ identically in the fits reported in Table~\ref{Tab:Bond_Density}.

From the fits, we estimate $\rhobranchLim = 0.214\,050\,18(5)$, $\rhojunctionLim = 0.035\,949\,79(8)$ and $\rhononbridgeLim = 0.250\,000\,1(2)$.
We note that $\rhobranchLim+\rhojunctionLim = \rhononbridgeLim = 1/4$ within error bars, as expected. 
The fit details are summarized in Table~\ref{Tab:Bond_Density}.
We also note that the estimates of $a_1$ for $\rhojunction$ and $\rhononbridge$ are equal in magnitude and opposite in sign, which is as expected given that $a_1$ is consistent with zero for $\rhobranch$.

\begin{figure}
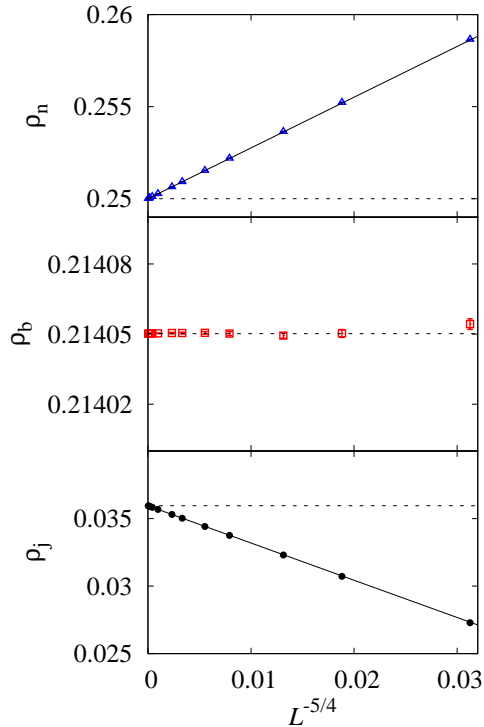

  \densitiesFig
  \caption{Plots of $\rhononbridge$ (top), $\rhobranch$ (middle), and $\rhojunction$ (bottom) versus $L^{-5/4}$.
    From top to bottom, the three dashed lines respectively correspond to values $1/4$, $0.214\,050\,18$, and $0.035\,949\,79$.
    The statistical error of each data point is smaller than the symbol size. The straight lines are simply to guide the eye.}
  \label{Fig:Bond_Density}
\end{figure}

In Fig.~\ref{Fig:Bond_Density}, we plot $\rhobranch$, $\rhojunction$ and $\rhononbridge$ versus $L^{-5/4}$.
The plot clearly demonstrates that the leading finite-size corrections for $\rhojunction$ and $\rhononbridge$ are governed by exponent $x_2=5/4$, 
while essentially no finite-size dependence can be observed for $\rhobranch$.

\begin{table}[htbp]
  \begin{tabular}[t]{|l|llll|}
    \hline
    $\rho$      &  $\rho_{0}$        & $y_1$        &  $a_1$         & $L_{\rm min}/DF/\chi^2$ \\
    \hline
        {\multirow{3}{*}{$\rhobranch$}}
        &  0.214\,050\,19(3) &$5/4$        & ~~0.000\,04(4)     &  24/9/6 \\
        &  0.214\,050\,19(3) &$5/4$        & ~~0.000\,05(5)     &  32/8/6 \\
        &  0.214\,050\,18(3) &$5/4$        & ~~0.000\,09(6)     &  48/7/4 \\
        \hline 
        
        {\multirow{3}{*}{$\rhojunction$}}
        &  0.035\,949\,78(5) &$1.250\,2(2)$&  $-0.277\,7(2)$     &  24/8/4 \\
        &  0.035\,949\,78(5) &$1.250\,2(3)$&  $-0.277\,7(3)$     &  32/7/4 \\
        &  0.035\,949\,79(6) &$1.250\,0(4)$&  $-0.277\,5(4)$     &  48/6/4 \\
        \hline
            {\multirow{3}{*}{$\rhononbridge$}}  
            &  0.250\,000\,1(1)  &$1.250\,7(5)$&  ~~0.278\,3(5)     &  24/8/2 \\
            &  0.250\,000\,1(1)  &$1.250\,8(6)$&  ~~0.278\,4(6)     &  32/7/2 \\
            &  0.250\,000\,1(1)  &$1.250\,6(7)$&  ~~0.278\,1(9)     &  48/6/2 \\
            \hline
  \end{tabular}
  \caption{Fit results for $\rhobranch$, $\rhojunction$, and $\rhononbridge$.}
  \label{Tab:Bond_Density}
\end{table}

\subsection{Fractal dimensions of clusters}
\label{Fractal dimension of clusters}
The first question to be addressed in this section is to determine if the fractal dimension of leaf-free clusters differs from $\dF$.
We therefore fit the data for $\leafFreeCluster$ to the ansatz~\eqref{Eq:Clusters_Loops}.
The fit results are reported in Table~\ref{Tab:FD_Clusters}.
In the reported fits we set $c_0=0$ identically, since leaving it free produced estimates for it consistent with zero.
Leaving $y_1$ free, we estimate $y_1 = 1.3(3)$, which is consistent with the value $y_1=5/4$ observed for $\rhojunction$ and $\rhononbridge$.

\begin{figure}
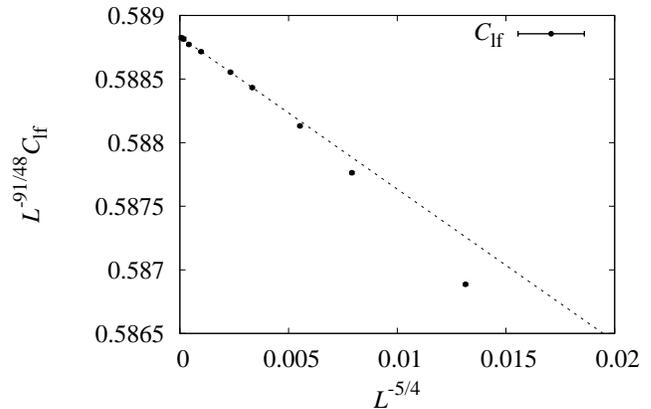

  \leafFreeFig
  \caption{Plot of $L^{-91/48}\leafFreeCluster$ versus $L^{-5/4}$. 
    The statistical error of each data point is smaller than the symbol size. The straight lines are simply to guide the eye.}
  \label{Fig:leaf-free}
\end{figure}

From the fits, we estimate $d_{\leafFreeCluster} = 1.895\,84(6)$, which is consistent with the fractal dimension of percolation clusters, $\dF=91/48$.
This indicates that although around $43\%$ of all occupied bonds are branches (see Table~\ref{Tab:Bond_Density}),
their deletion from percolation configurations does not alter the fractal dimension of the resulting clusters.
In Fig.~\ref{Fig:leaf-free}, we plot $L^{-91/48}\leafFreeCluster$ versus $L^{-5/4}$.

For comparison, we also performed fits of $\percolationCluster$ to the ansatz~\eqref{Eq:Clusters_Loops},
obtaining the estimate $a_0=0.983\,8(5)$, which is strictly larger than the value estimated for $\leafFreeCluster$. 
As $L\to\infty$ therefore, a non-trivial fraction $1-a_0(\leafFreeCluster)/a_0(\percolationCluster)\approx 40\%$ of sites in the largest percolation cluster are deleted by burning the branches.
This is close to, but slightly smaller than, the proportion of occupied bonds which are branches $2\rhobranch\approx 43\%$.

We next study the fractal dimension of bridge-free clusters.
We fit the Monte Carlo data for $\bridgeFreeCluster$ to the ansatz~\eqref{Eq:Clusters_Loops}, and the results are reported in Table~\ref{Tab:FD_Clusters}.
In the fits, we fixed $y_1=5/4$, and again observed that $c_0$ is consistent with zero.
We also performed fits (not shown) with $y_1$ free, or fixed to $y_1=1$, in order to estimate the systematic error in our estimates of $\dB$.
This produced our final estimate $\dB=1.643\,36(10)$. 
This value is consistent with the estimate $\dB=1.643\,4(2)$~\cite{DengBloteNienhuis04a}, but with an improved error bar.

\begin{figure}
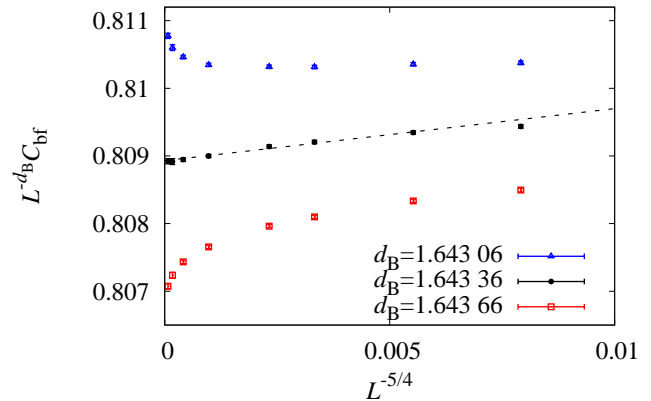

  \backboneFig
  \caption{Plot of $L^{-\dB}\bridgeFreeCluster$ versus $L^{-5/4}$, with $\dB = 1.643\,06$, 1.643\,36 and 1.643\,66.
    The statistical error of each data point is smaller than the symbol size. The straight lines are simply to guide the eye.}
  \label{Fig:backbone}
\end{figure}

Fig.~\ref{Fig:backbone} plots $L^{-\dB}\bridgeFreeCluster$ versus $L^{-5/4}$, with $\dB$ chosen to be the central value of our estimate, as well as the central value plus or minus three error bars.
The obvious upward (downward) bending as $L$ increases when using a $\dB$ value above (below) our central estimate illustrates the reliability of our final estimate of $\dB$.

\begin{table}[htbp]
  \begin{tabular}[t]{|l|lllll|}
    \hline
    $\scrO$ & $d_{\scrO}$  &  $a_0$        &  $a_1$     & $a_2$         & $L_{\rm min}/DF/\chi^2$ \\
    \hline
        {\multirow{3}{*}{$\leafFreeCluster$}}
        &$1.895\,82(2)$&  0.588\,88(2) &$-0.103(6)$ & $-0.61(5)$     &  24/7/8 \\
        &$1.895\,84(2)$&  0.588\,81(6) &$-0.091(9)$ & $-0.75(9)$   &  32/6/4 \\
        &$1.895\,84(2)$&  0.588\,78(8) &$-0.08(2)$  & $-0.8(3)$      &  48/5/4 \\
        \hline
            {\multirow{3}{*}{$\bridgeFreeCluster$}}
            & $1.643\,32(3)$   &  0.809\,2(2)  &  ~~0.07(2) &  $-0.2(2)$    &  24/7/4 \\
            & $1.643\,32(3)$   &  0.809\,1(2)  &  ~~0.08(3) &  $-0.3(3)$    &  32/6/4 \\
            & $1.643\,36(4)$   &  0.808\,9(3)  &  ~~0.14(5) &  $-1.2(7)$    &  48/5/1 \\
            \hline
  \end{tabular}
  \caption{Fit results for $\leafFreeCluster$ and $\bridgeFreeCluster$.}
  \label{Tab:FD_Clusters}
\end{table}

\subsection{Fractal dimensions of loops}
\label{Fractal dimension of loops}
Finally, we studied the fractal dimensions of the loop configurations associated with both leaf-free and bridge-free configurations.

We fit the data for $\leafFreeHull$ and $\bridgeFreeHull$ to the ansatz~\eqref{Eq:Clusters_Loops}, with $y_1=5/4$ fixed.
For both $\leafFreeHull$ and $\bridgeFreeHull$, the fits gave estimates of $c_0$ consistent with zero. 
We therefore fixed $c_0=0$ identically in the fits reported in Table~\ref{Tab:FD_Loops}.
To estimate the systematic error, we compared these results with fits in which $y_1$ was free, and also fits with $y_1=1$ fixed.
Our resulting final estimates are $d_{\leafFreeHull} = 1.749\,96(8)$ and $d_{\bridgeFreeHull} = 1.333\,3(3)$.

For leaf-free configurations, therefore, our fits strongly suggest $d_{\leafFreeHull} = 7/4 = \dH$.
Thus, deleting branches from percolation configurations affects neither the fractal dimension for cluster size, nor the fractal dimension for lengths of the associated loops.
For bridge-free configurations by contrast, the fits suggest that $d_{\bridgeFreeHull} = 4/3 = \dE$.

\begin{figure}
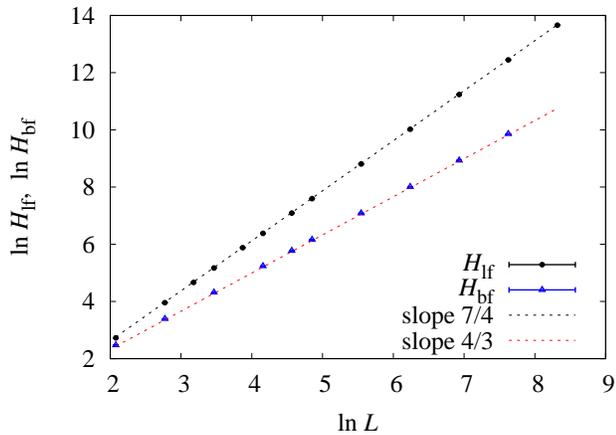

  \hullFig
  \caption{Log-log plot of $\leafFreeHull$ and $\bridgeFreeHull$ versus $L$. 
    The two dashed lines have slopes 7/4 and 4/3 respectively. 
    The statistical error of each data point is smaller than the symbol size.}
  \label{Fig:FD_loops}
\end{figure}

In Fig.~\ref{Fig:FD_loops}, we plot $\leafFreeHull$ and $\bridgeFreeHull$ versus $L$ to illustrate our estimates for $d_{\leafFreeHull}$ and $d_{\bridgeFreeHull}$.

\begin{table}[htbp]
  \begin{tabular}[t]{|l|lllll|}
    \hline
    $\scrO$ & $d_{\scrO}$         &  $a_0$        &  $a_1$     & $a_2$         & $L_{\rm min}/DF/\chi^2$ \\
    \hline
        {\multirow{3}{*}{$\leafFreeHull$}}
        &$1.750\,05(2)$ &  0.408\,11(6)  & $0.039(6)$ & $-0.25(5)$    &  24/7/12 \\
        &$1.750\,02(3)$ &  0.408\,17(7)  & $0.029(9)$ & $-0.15(9)$    &  32/6/10 \\
        &$1.749\,99(4)$ &  0.408\,30(9)  & $0.00(2)$  & ~~$0.3(3)$    &  48/5/5 \\
        \hline
            {\multirow{3}{*}{$\bridgeFreeHull$}}
            &$1.333\,33(8)$ &  0.734\,0(4)  & $0.28(3)$  &  $-1.1(2)$    &  16/5/4 \\
            &$1.333\,2(2)$  &  0.734\,5(6)  & $0.20(8)$  &  $-0.3(8)$    &  32/4/3 \\
            &$1.333\,3(2)$  &  0.734\,2(9)  & $0.3(2)$   &  $-2(3)$      &  64/3/2 \\
            \hline 
  \end{tabular}
  \caption{Fit results for $\leafFreeHull$ and $\bridgeFreeHull$.}
  \label{Tab:FD_Loops}
\end{table}

\section{Discussion}
\label{Discussion}
We have studied the geometric structure of percolation on the torus, by considering a partition of the edges into three natural classes.
On the square lattice, we have found that leaf-free configurations have the same fractal dimension and hull dimension as standard percolation configurations,
while bridge-free configurations have cluster and hull fractal dimensions consistent with the backbone and external perimeter dimensions, respectively.

In addition to the results discussed above, we have extended our study of leaf-free configurations to site percolation on the triangular lattice and bond percolation on the simple-cubic lattice,
the critical points of which are respectively 1/2 and 0.248\,811\,82(10)~\cite{WangZhouZhangGaroniDeng13}.
We find numerically that the fractal dimensions of leaf-free clusters for these two models are respectively 1.895\,7(2) and 2.522\,7(6),
both of which are again consistent with the known results $91/48$ and $2.522\,95(15)$~\cite{WangZhouZhangGaroniDeng13} for $\dF$.
In both cases, our data show that the density of branches is again only very weakly dependent on the system size.

It would also be of interest to study the bridge-free configurations on these lattices.
In addition to investigating the fractal dimensions for cluster size, and in the triangular case also the hull length,
it would be of interest to determine whether the leading finite-size correction to $\rhononbridge$ is again governed by the two-arm exponent.

The two-arm exponent is usually defined by considering the probability of having multiple spanning clusters joining inner and outer annuli in the plane.
As noted in Section~\ref{Fitting methodology} however, our results show that for percolation, the two-arm exponent also governs the probability of a rather natural geometric event on the torus:
the event that a given edge is not a bridge but has both its loop arcs in the same loop.
This provides an interesting alternative interpretation of the two-arm exponent in terms of toroidal geometry.

Let us refer to an edge that is not a bridge but has both its loop arcs in the same loop as a {\em pseudobridge}.
We note that an alternative interpretation of the observation that $(\rhononbridge-\rho_2)\sim L^{-x_2}$ is that the number of pseudobridges $L^2(\rhononbridge-\rho_2)$ scales as $L^{\dR}$.

A natural question to ask is to what extent the above results carry over to the general setting of the Fortuin-Kasteleyn random-cluster model. Consider the case of two dimensions once more.
In that case, we know that if we fix the edge weight to its critical value and take $q\to0$ we obtain the uniform spanning trees (UST) model.
For this model all edges are branches, and so the leaf-free configurations, which are therefore empty, certainly do not scale in the same way as UST configurations.
Despite this observation, 
preliminary simulations~\footnote{These simulations were performed using the Sweeny algorithm~\cite{Sweeny82} for $q<1$ and the Chayes-Machta algorithm~\cite{ChayesMachta98} for $q>1$.}
performed on the toroidal square lattice at $q=0.09$, $0.16$, $1.5$, $2.0$, $2.5$, $3.0$ and $3.5$ suggest that, for all $q\in(0,4]$,
the leaf-free configurations have the same fractal dimension and hull dimension as the corresponding standard random cluster configurations.
In the context of the random cluster model, the behaviour of the leaf-free configurations for the UST model therefore presumably arises via amplitudes which vanish at $q=0$.

In addition, these preliminary simulations suggest that the number of pseudobridges in fact scales as $L^{\dR}$ for the critical random cluster model at any $q\in(0,4]$.
It would also be of interest to determine whether the fractal dimensions of cluster size and hull length for bridge-free random cluster configurations again coincide with $\dB$ and $\dE$ when $q\neq 1$.

\section{Acknowledgments}
\label{Acknowledgments}
The authors wish to thank Bob Ziff for several useful comments/suggestions, and T.G. wishes to thank Nick Wormald for fruitful discussions relating to the density of bridges.
This work is supported by the National Nature Science Foundation of China under Grant No. 91024026 and 11275185, and the Chinese Academy of Sciences.
It was also supported under the Australian Research Council's Discovery Projects funding scheme (project number DP110101141), 
and T.G. is the recipient of an Australian Research Council Future Fellowship (project number FT100100494).
The simulations were carried out in part on NYU's ITS cluster, which is partly supported by NSF Grant No. PHY-0424082.
In addition, this research was undertaken with the assistance of resources provided at the NCI National Facility through the National Computational Merit Allocation Scheme supported by the Australian Government.
J.F.W and Y.J.D acknowledge the Specialized Research Fund for the Doctoral Program of Higher Education under Grant No. 20103402110053.
Y.J.D also acknowledge the Fundamental Research Funds for the Central Universities under Grant No. 2340000034.

\appendix

\section{A loop duality lemma}
\label{loop lemma appendix}
Let $\scrL_1$ ($\scrL_2$) denote the fraction of occupied edges whose two associated loop segments belong to the same (distinct) loop(s).

\begin{figure}[t]
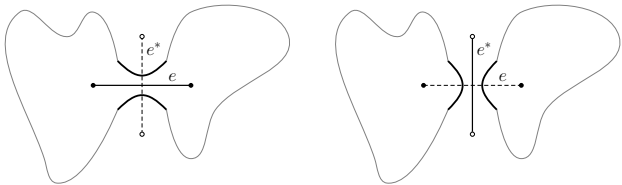

  \begin{center}
    \loopFigA
    \loopFigB
  \end{center}
  \caption{
    Left: Illustration of a configuration $A\subseteq E$ for which the event $\ell_1(e)$ occurs.
    Right: The corresponding configuration $A^*\cup e^*$ for which the event $\ell_2(e^*)$ occurs.
  }
  \label{Fig:loops diagram}
\end{figure}

\begin{lemma}
  \label{loop lemma}
  Consider $p=1/2$ bond percolation on $\ZZ_L^2$. For any $L$ we have $\EE\,\scrL_1 = \EE\,\scrL_2 = 1/4$.
  \begin{proof}
    Let $m=2L^2$ denote the number of edges in $G=\ZZ_L^2$. 
    Since $G$ is a cellularly-embedded graph~\cite{EllisMonaghanMoffatt13}, it has a well-defined geometric dual $G^*$ and medial graph $\scrM(G)=\scrM(G^*)$.
    For any $e\in E$ we denote its dual by $e^*\in E^*$.
    
    For $e\in E$, let $\ell_1(e)$ be the event that the two loop segments associated with $e$ both belong to the same loop, and let $\ell_2(e)$ be the event that they belong to distinct loops.
    The key observation is that for any $0\le a \le m$ we have
    \begin{equation}
      \sum_{\ontop{A\subseteq E}{|A|=a}} \sum_{e\in A} \ind_{\ell_1(e)}(A)
      =
      \sum_{\ontop{B^*\subseteq E^*}{|B^*|=m+1-a}} \sum_{e^*\in B^*} \ind_{\ell_2(e^*)}(B^*)
      \label{loops identity}
    \end{equation}
    To see this, first note that the number of terms on either side of~\eqref{loops identity} is $\binom{m}{a}a = \binom{m}{m+1-a}(m+1-a)$, and that each term is either 0 or 1.
    Then note that there is a bijection between the terms on the left- and right-hand sides such that the term on the left-hand side is 1 iff the term on the right-hand side is 1, as we now describe.
    Let $A\subseteq E$ with $|A|=a$, and let $A^*$ denote the {\em dual} configuration: include $e^*$ in $A^*$ iff $e\not\in A$.
    With the term on the left-hand side corresponding to $(A,e)$, associate the term $(B^*,e^*)=(A^*\cup e^*,e^*)$ appearing on the right-hand side. This is clearly a 1-1 correspondence.

    Let $\fL(A)$ denote the loop configuration corresponding to $A$.
    By construction, $\fL(A)=\fL(A^*)$. The loop configuration $\fL(A^*\cup e^*)$ differs from $\fL(A)$ only in that the loop arcs cross $e^*$ in $\fL(A)$ but cross $e$ in $\fL(A^*\cup e^*)$.
    If $\ind_{\ell_1(e)}(A) = 1$, then it follows that $\ind_{\ell_2(e^*)}(B^*)=1$.
    The converse holds by duality, and so~\eqref{loops identity} is established. See Fig.~\ref{Fig:loops diagram} for an illustration.

    Summing both sides of~\eqref{loops identity} over $a$ and dividing by $m\,2^{m}$ then shows that $\EE\,\scrL_1 = \EE\,\scrL_2$.
    Since on average precisely 1/2 of all edges are occupied when $p=1/2$, the stated result follows.
  \end{proof}
\end{lemma}

\end{document}